\documentclass[reprint, amsmath, amssymb, aps, prl]{revtex4-2}

\usepackage{graphicx}% Include figure files
\usepackage{dcolumn}% Align table columns on decimal point
\usepackage{bm}% bold math
\usepackage{braket}
\usepackage{esvect}
\usepackage[dvipsnames]{xcolor}

\usepackage{xr-hyper}
\usepackage{hyperref}

\usepackage[margin=0.8in]{geometry}
\usepackage{physics}
\usepackage{amsmath}
\usepackage{mathtools}
\usepackage[framed,numbered]{matlab-prettifier}
\usepackage{hyperref}
\usepackage{newfloat}
\DeclareFloatingEnvironment[name={Figure}]{suppfigure}

\usepackage[normalem]{ulem}

\renewcommand{\vec}[1]{\mathbf{#1}}

\usepackage[dvipsnames]{xcolor}
\usepackage{soul}
 
\newcommand{\qutech}{QuTech and Kavli Institute of Nanoscience, Delft University of Technology, 2600 GA Delft, The Netherlands}

\newcommand{\tue}{Department of Applied Physics, Eindhoven University of Technology, MB Eindhoven, 5600, The Netherlands}
\newcommand{\uiuc}{Department of Physics and Frederick Seitz Materials Research Laboratory, University of Illinois at Urbana-Champaign, Urbana, IL 61801, USA}
\begin{document}
\preprint{APS/123-QED}
\title{
Variable and orbital-dependent spin-orbit field orientations in a InSb double quantum dot characterized via dispersive gate sensing
}

\author{Lin~Han$^{1}$} 
    \email{L.Han-2@tudelft.nl}
\author{Michael~Chan$^{1}$}
\author{Damaz~de~Jong$^{1}$}
\author{Christian~Prosko$^{1}$}
\author{Ghada~Badawy$^{2}$}
\author{Sasa~Gazibegovic$^{2}$}
\author{Erik~P.A.M.~Bakkers$^{2}$}
\author{Leo~P.~Kouwenhoven$^{1}$}
\author{Filip~K.~Malinowski$^{1}$}
    \email{F.K.Malinowski@tudelft.nl}
\author{Wolfgang~Pfaff$^{3}$}
\affiliation{$^1$\qutech\\$^2$\tue\\$^3$\uiuc}

\begin{abstract}
Utilizing dispersive gate sensing (DGS), we investigate the spin-orbit field ($\vec{B}_{SO}$) orientation in a many-electron double quantum dot (DQD) defined in an InSb nanowire.
While characterizing the inter-dot tunnel couplings, the measured dispersive signal depends on the electron charge occupancy, as well as on the amplitude and orientation of the external magnetic field.
The dispersive signal is mostly insensitive to the external field orientation when a DQD is occupied by a total odd number of electrons.
For a DQD occupied by a total even number of electrons, the dispersive signal is reduced when the finite external magnetic field aligns with the effective $\vec{B}_{SO}$ orientation.
This fact enables the identification of $\vec{B}_{SO}$ orientations for different DQD electron occupancies.
The $\vec{B}_{SO}$ orientation varies drastically between charge transitions, and is generally neither perpendicular to the nanowire nor in the chip plane.
Moreover, $\vec{B}_{SO}$ is similar for pairs of transitions involving the same valence orbital, and varies between such pairs.
Our work demonstrates the practicality of DGS in characterizing spin-orbit interactions in quantum dot systems, without requiring any current flow through the device.
\end{abstract}

\maketitle

% ------------------- CONTENT -------------------

% ================== 1. Introduction ==================
%-----
% Figure-1-start
%-----
\begin{figure}[tbh]
\includegraphics[scale=0.68]{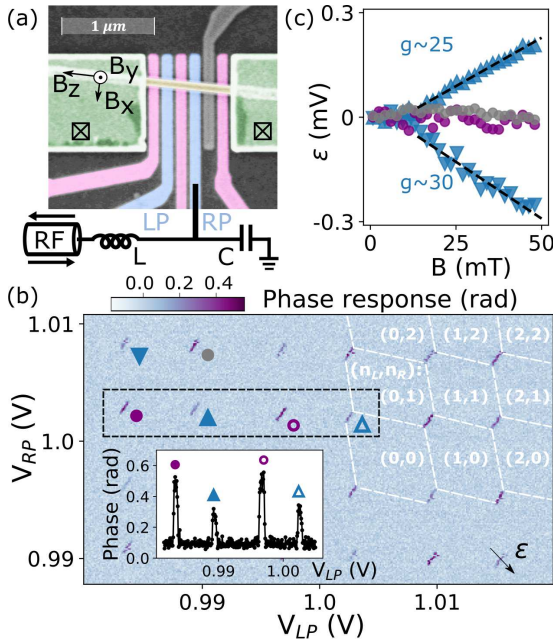}
\caption{ \label{fig:Fig1} 
(a) False-colored SEM image of the device and the circuit schematics for DGS.
The barrier and plunger gates are indicated in pink and blue respectively, while the metallic leads are indicated in green. 
The grey gate is electrostatically floating (unused in this work).
(b) Charge stability diagram of the DQD at zero magnetic field.
The inset shows in the range marked by the dashed black rectangular, the maximum phase response for each $V_{LP}$.
(c) The position of four neighboring ICTs along detuning axis, as a function of arbitrarily oriented external field. 
Markers in (b) indicate the relative charge occupancy of the ICTs, although they do not correspond to the same ICTs as in (c). 
}
\end{figure}
%-----
% Figure-1-end
%-----
%%% 
%%% Introduction
%%%
        % 1- what is SOI?
A spinful carrier moving in an electromagnetic field may experience a coupling between its spin and momentum degree of freedom, namely spin-orbit interaction (SOI). 
        % 1- uses of SOI.
The SOI allows spin manipulation with electric fields in semiconductor platforms, such that it enables electric dipole spin resonance \cite{rashba2003orbital, vzutic2004spintronics, golovach2006electric, hanson2007spins}, spin-cavity couplings \cite{nadj2010spin, petersson2012circuit, tarucha2016spin, crippa2019gate}, while it also enhances effects detrimental to spin-based quantum information processing: relaxation and decoherence \cite{kloeffel2013prospects, chatterjee2021semiconductor}.
        % 1- what is Bso?
For many cases, SOI can be described as an effective spin-orbit field ($\vec{B}_{SO}$) acting on the carriers.
Notably, $\vec{B}_{SO}$ associated with the Rashba SOI is perpendicular to both the electric field $\vec{E}$ and the carrier momentum $\vec{p}$, following $\vec{B}_{SO}\propto\vec{E}\times\vec{p}$ \cite{winkler2003spin, ihn2009semiconductor}.
        % 1- uses of Bso.
In ideal nanowire systems, carriers are confined in a one-dimensional path, which forces their momentum $\vec{p}$ to be along the nanowire.
With the application of bottom electrostatic gates, the assumed electric field $\vec{E}$ is perpendicular to the substrate surface.
Accordingly, the $\vec{B}_{SO}$ orientation is expected and has been experimentally proved to be not only in-plane of the chip, but also nearly perpendicular to a bottom-gated nanowire \cite{ihn2009semiconductor, kammhuber2017conductance}.
Despite the electrostatic confinement, this conclusion is further found to hold for electron tunneling in few-electron double quantum dots (DQD), even when the center-to-center distance between the dots is small with respect to typical spin orbit lengths \cite{nadj2012spectroscopy, wang2018anisotropic, dorsch2019side}.
        % 1- uses of Bso of in MZM research.
Knowing the $\vec{B}_{SO}$ orientation in such nanowires is particularly important for semiconductor-superconductor hybrid systems that aim to realize Majorana zero modes, as setting the external magnetic field perpendicular to $\vec{B}_{SO}$ is a precondition to open a topological gap \cite{lutchyn2018majorana}.

        % 2- old tech for Bso
The conventional way to characterize SOI is associated with tunneling between quantum dots, which employs bias voltages across a DQD segment and measurements of spin blockade leakage current \cite{pfund2007spin,nadj2010disentangling,nadj2012spectroscopy,bogan2021spin}. 
However, scalable qubit devices \cite{zajac2016scalable, karzig2017scalable} may favor characterization methods that do not require transport measurements.
        % 2- our tech for Bso
Here, we explore dispersive gate sensing (DGS) \cite{colless2013dispersive, crippa2019gate, zheng2019rapid, de2019rapid, van2019revealing, sabonis2019dispersive, de2020dispersive} to characterize SOI, especially the $\vec{B}_{SO}$ orientation.  
Our protocol does not employ transport measurements, is compatible with fast data acquisition in rastering schemes \cite{stehlik2015fast, de2019rapid}, and is promising for the integration of qubit characterization and readout capabilities \cite{plugge2017majorana, smith2020dispersive}.

% ================== 2. Setup ==================          

        % 3- device // fig1a
The device under study is depicted in Fig.~\ref{fig:Fig1}(a). 
An InSb nanowire is placed on top of prefabricated bottom finger gates.
The barrier gates confine the electrons and control the tunnel coupling within the DQD, while the plunger gates LP (RP) tune the chemical potential of the left (right) dot.
The nanowire is grown along [111] direction, such that the Rashba SOI is expected to be dominant \cite{nadj2012spectroscopy, kammhuber2017conductance, badawy2019high}
        % 3- resonator
To implement DGS, the RP gate is coupled to an off-chip superconducting spiral-inductor resonator, with resonance frequency $f_{0}\approx$ 318.4 MHz and nominal inductance $L=$ 730 nH \cite{hornibrook2014frequency}.
        % 3- Cq (~df ~A&P)
At interdot charge transitions (ICTs), where the chemical potential for an electron residing in the left and the right dot are equal, the hybridization of electron wave functions between the two dots leads to an additional quantum capacitance $C_q$ loading the resonator, which is observable as a shift of $f_{0}$ \cite{duty2005observation,petersson2010charge,colless2013dispersive,de2019rapid}.
While fixing the probing frequency and detecting the reflected signal from the resonator, $f_{0}$ is translated into a change of reflection coefficient, thus to the amplitude and phase response.
        % 3- fitting method
In this letter, we fit the measured reflection coefficient with an analytical resonator model to extract $f_{0}$ and $C_q$ (see Suppl. A).
        % 3- temperature
All measurements were performed in a dilution refrigerator at a base temperature $T\approx$ 30~mK.
  
        % 4- CSD
In Fig.~\ref{fig:Fig1}(b), the charge stability diagram of the DQD is mapped by measuring the reflected phase response versus gate voltages $V_{LP}$ and $V_{RP}$. 
        % 4- ICT
It reveals a grid of ICTs with the lead-to-dot transitions hardly visible (marked by white lines), since the outer barrier gates are nearly pinched off.
        % 4- spacing/phase
Along $V_{LP}$ and $V_{RP}$ axes, both the spacings between the ICTs and the measured phase shifts at the ICTs tend to alternate between smaller and larger values (inset of Fig.~\ref{fig:Fig1}(b)).
        % 4- explain (is it clear now?)
As loading every other additional electron requires compensating for the level spacing on top of the charging energy, the smaller spacings along $V_{LP}$ ($V_{RP}$) are associated with having an odd number of electrons in the left (right) dot \cite{van2002electron}.
Furthermore, an ICT corresponding to a total odd number of electrons in the DQD exhibit a larger phase shift, since the spin degeneracy of having total even charges leads to a reduction of the maximum of $C_q$ \cite{hanson2007spins, cottet2011mesoscopic, van2019revealing}.

        % 4- prof by B
The identification of the total charge parity in the DQD is additionally verified by applying an external magnetic field $\vec{B}$ (Fig.~\ref{fig:Fig1}(c)) \cite{schroer2012radio}.
For four neighboring ICTs, their positions along the detuning axis are measured as a function of $\vec{B}$, with detuning $\varepsilon \coloneqq [(V_{LP}-V_{LP, \varepsilon=0})-(V_{RP}-V_{RP, \varepsilon=0})]/2$.
%\equiv 
We observe shifts only for the ICTs with a total even occupancy, consistent with Zeeman effect.
Fits to the data for even-occupied ICTs in a region exhibiting a linear shift in magnetic field \cite{hanson2007spins} yield the effective g-factors of approximately 25 and 30.
        % 4- labels
Based on these observations, the parity of the electron numbers in the DQD is indicated with labels $(n_L,n_R)$, with $n_{L(R)}$ indicating the excess number of electrons with respect to an even number of electrons $(N_L, N_R)$ on the left (right) dot (also indicated in Fig.~\ref{fig:Fig1}(b)).
The number of electrons in each dot is estimated to be in the range of 70 to 150 electrons, considering the plunger gate voltages, pinch off voltages, and the spacing between ICTs.

\begin{figure}[tbh]
\includegraphics[scale=0.68]{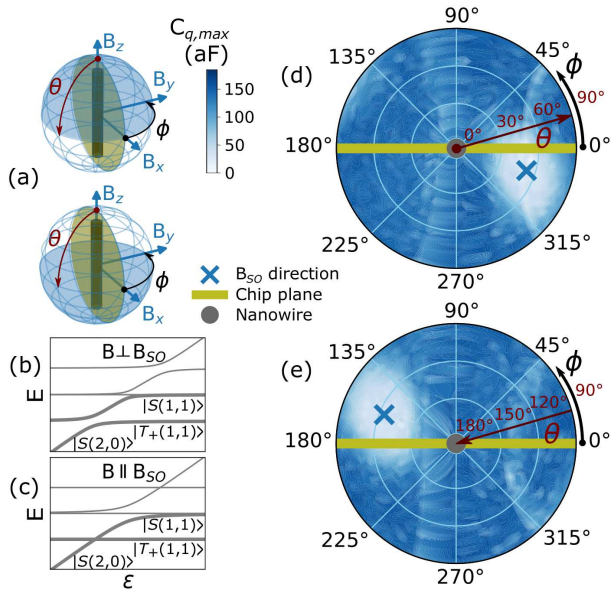}
\caption{\label{fig:Fig2} 
(a) Illustration of spherical coordinates with respect to the nanowire and the substrate.
The top and bottom panel correspond to (d) and (e) respectively.
(b-c) Schematic energy diagrams of a DQD for (b)~$\vec{B}\perp\vec{B}_{SO}$ and (c)~$\vec{B}\parallel\vec{B}_{SO}$.
The (avoided) crossing between the two lowest states are highlighted.
$C_{q,max} = 0$ when they cross, due to the flat shape of $\ket{T_{+}(1,1)}$.
(d-e) The extracted $C_{q,max}$ values plotted on an external magnetic field angle map.
Polar projection of the map is shown in
(d)~when $0^\circ \leq \theta \leq 90^\circ$, and
(e)~when $90^\circ \leq \theta \leq 180^\circ$.
The blue crosses mark the characterized $\pm\vec{B}_{SO}$ orientations, which is the center of the region where $C_{q,max}$ is suppressed.}
\end{figure}

% ================== 4. Identify Bso field ==================
        % 5- focus on even ICT
Having identified the total charge parity of the ICTs, we characterize the $\vec{B}_{SO}$ field orientation for an even-occupied ICT.
        % 5- how to get data // fig2a
We apply an external magnetic field with fixed amplitude $|\vec{B}| = 30$~mT.
$C_{q,max}$ which denotes the maximum values of $C_q$ at the ICT (see Suppl. A) is extracted as a function of the field orientation in spherical coordinates $\varphi$ and $\theta$ (Fig.~\ref{fig:Fig2}(a)).
        % 5- how to plot data // fig2de
Fig.~\ref{fig:Fig2} (d) and (e) display the obtained data in range $0^\circ \leq \theta \leq 90^\circ$, and $90^\circ \leq \theta \leq 180^\circ$, respectively.
        % 5- what is in data
There are two regions at which $C_{q,max}$ is strongly suppressed. 
They lie at opposite directions in the spherical coordinates, neither perpendicular to the nanowire, nor in plane of the substrate. 
We interpret the centers of the suppression regions as corresponding to the directions parallel and anti-parallel to $\vec{B}_{SO}$.
Energy diagrams of the DQD are presented in Fig.~\ref{fig:Fig2}(b,c).
        % 6- energy diagram // fig2bc
        % 6- fig2b
Due to different total spin, the lowest singlet state $\ket{S(2,0)}$ and triplet state $\ket{T_{+}(1,1)}$ only couple if electron spin flips during tunneling are allowed.
This coupling arises only when $\vec{B}$ is not aligned with $\vec{B}_{SO}$ (Fig.~\ref{fig:Fig2}(b)), resulting in finite curvature of the ground state energy at the ICT, and thus finite $C_{q,max}$.
        % 6- fig2c
When $\vec{B} \parallel \vec{B}_{SO}$, the two states do not couple (Fig.~\ref{fig:Fig2}(c)), therefore $C_{q,max}$ is suppressed because of the flat energy dispersion of $\ket{T_{+}(1,1)}$ state \cite{hanson2007spins, mizuta2017quantum, esterli2019small}.

        % 7- unexpected Bso
The observation that $\vec{B}_{SO}$ is neither perpendicular to the nanowire nor in-plane of the chip can be attributed to several reasons.
First, the complicated gate structure is likely to create a nonuniform potential, making the local electric fields deviate significantly from the out-of-plane direction.
Second, staying in many-electron regime bring more complexity, as the overlap between the wave functions of the two dots may not spatially coincide with the direction of the nanowire.
Consequently, the momentum associated with electron tunneling is not necessarily along the nanowire.
Third, although not dominant, a finite contribution of Dresselhaus SOI may also exist, so that spin modulations in the cross-sectional plane contribute to the offset angle with respect to the chip plane \cite{bringer2019dresselhaus}.

%-------------------------------------------------Figure-2: start
%-------------------------------------------------Figure-2: end
	
%-------------------------------------------------Figure-3: start
\begin{figure*}[tbh]
\includegraphics[scale=0.68]{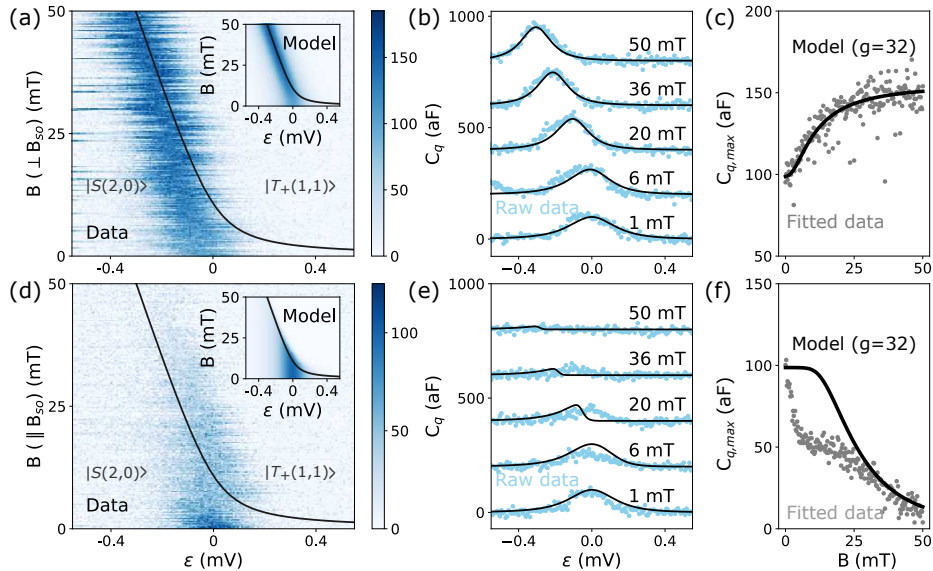}
\caption{\label{fig:Fig3} 
Evolution of the ICT in external magnetic field $\vec{B}$ for (a-c) $\vec{B}\perp\vec{B}_{SO}$, and (d-f) $\vec{B}\parallel\vec{B}_{SO}$.
(a,d) $C_q$ as a function of $\vec{B}$ and $\varepsilon$.
The inset presents the numerical simulation.
Black curves in both the main figure and the inset indicate the degeneracy point between $\ket{S(2,0)}$ and $\ket{T_{+}(1,1)}$.
(b,e) Line cuts of (a,d) for several magnitudes of $\vec{B}$.
For clarity, they are separated in the $C_q$ axis by 200 aF.
Black curves come from the simulation.
(c,f) $C_{q,max}$ as a function of $\vec{B}$.
Grey dots are extracted from a phenomenological Gaussian fit of the data. 
Black curves indicate the $C_{q,max}$ taken from the insets of (a,d).
}
\end{figure*}
%-------------------------------------------------Figure-3: end	
	
%================== 5. Field dependence of Cq ==================
        % 8- Cq vs B
Next, we study the evolution of $C_q$ at the same ICT, as a function of $\vec{B}$ and $\varepsilon$.
        % 8- perp. case
While increasing the amplitude of $\vec{B}\perp\vec{B}_{SO}$, we find a nearly linear shift of the $C_q$ maximum along detuning axis (Fig.~\ref{fig:Fig3}(a,b)). 
This is accompanied by a gradual increase of $C_{q,max}$ value (Fig.~\ref{fig:Fig3}(c)), starting at about $100$~aF when $\vec{B}$ is zero, and saturating near $150$~aF for $\vec{B}$ above 25~mT.
        % 8- para. case
In contrast, $C_{q,max}$ is suppressed (Fig.~\ref{fig:Fig3}(d-f)) for $\vec{B}\parallel \vec{B}_{SO}$, since $\vec{B}_{SO}$ no longer introduces singlet-triplet coupling in this orientation. 
Along $\vec{B}$, the suppression occurs in two distinct steps (see Fig.~\ref{fig:Fig3}(f)). 
Initially, $C_{q,max}$ drops rapidly from 100~aF for low $\vec{B}$, and starts saturating near the value of 50~aF with $\vec{B}$ above $\sim$10~mT. 
Then, $C_{q,max}$ starts dropping even further at about 25~mT. 
In the limited measurement range, $C_{q,max}$ appears to be trending towards zero.

        % 9- model
To understand the capacitative response of the ICT in magnetic field, we employ a two-site Hubbard model (Suppl. B) \cite{ziesche1997two}. 
The SOI in our model is phenomenologically described as an effective field which can point in an arbitrary direction in space, namely both Rashba and Dresselhaus SOI are taken into consideration.
The model includes the spin precessing tunneling matrix element $t_p$ as part of total tunneling strength $t_{tot}$.
The term $t_p$ depends on the SOI strength and modulates spin-flip together with an angle between $\vec{B}$ and $\vec{B}_{SO}$. 
The individual cuts at $B=0$ and $B=50$~mT for $\vec{B}\perp\vec{B}_{SO}$ are used to estimate $t_{tot}=20$~$\mu$eV and $t_p=18$~$\mu$eV, by fitting the analytical expressions of $C_q$ in Suppl. B. 
For simplicity, we assume isotropic g-factors $g=32$ being equal in both dots, with the value taken from linear shifts of charge transitions as in Fig.~\ref{fig:Fig3}(a). 
The effective lever arm of the gate attached to the resonator is $\alpha=0.26$, according to the ratio between the height and width of a Coulomb diamond, and an estimated crosstalk between the gates of $20\%$.
The electron temperature in the model is set to 30~mK, based on a Coulomb blockade thermometry measurement performed before this experiment.
        % 9- simulation
The simulated results are illustrated in the insets of Fig.~\ref{fig:Fig3}(a,d), and in black in Fig.~\ref{fig:Fig3}(b,c,e,f). 
For $\vec{B}\perp\vec{B}_{SO}$, we find an excellent agreement with the data in a full range of magnetic fields, with no free parameters. 
In contrast, for $\vec{B}\parallel\vec{B}_{SO}$, the simulated shift of the ICT along detuning axis is greater than observed when $\vec{B}$ is below 25~mT. 
Furthermore, the model does not qualitatively capture the two-stage suppression of $C_{q,max}$ when $\vec{B}$ increases (Fig.~\ref{fig:Fig3}(f)).

        % 10- explain discrepancy
We identify two elements in our model potentially responsible for the discrepancy. 
First, we consider possible g-factor non-uniformity and anisotropy \cite{nadj2012spectroscopy, qu2016quantized, mu2020measurements}. 
In particular, a smaller g-factor for $\vec{B}\parallel\vec{B}_{SO}$ can reduce the shift in detuning of the ICT in Fig.~\ref{fig:Fig3}(d), and eventually increase $\vec{B}$ at which the predicted suppression of $C_{q,max}$ occurs. 
Second, when $\vec{B}\parallel\vec{B}_{SO}$, spin relaxation rates mediated by hyperfine and SOI are hindered \cite{fujita2016signatures, lundberg2021non}.
As a consequence, Pauli spin blockade traps the system in one of the excited states which do not contribute to $C_{q,max}$.
        % 10- summary
We hypothesize that unaccounted Pauli spin blockade is responsible for the suppression of the $C_{q,max}$ in range of 0-25~mT. 
Meanwhile, the $C_{q,max}$ suppression above 25~mT is consistent with our model, except it occurs at higher field due to an overestimated g-factor.
This can be identified from comparing the $C_q$ peaks of model and data in Fig.~\ref{fig:Fig3}(e).

%-------------------------------------------------Figure-4: start
\begin{figure}[tbh]
\includegraphics[scale=0.68]{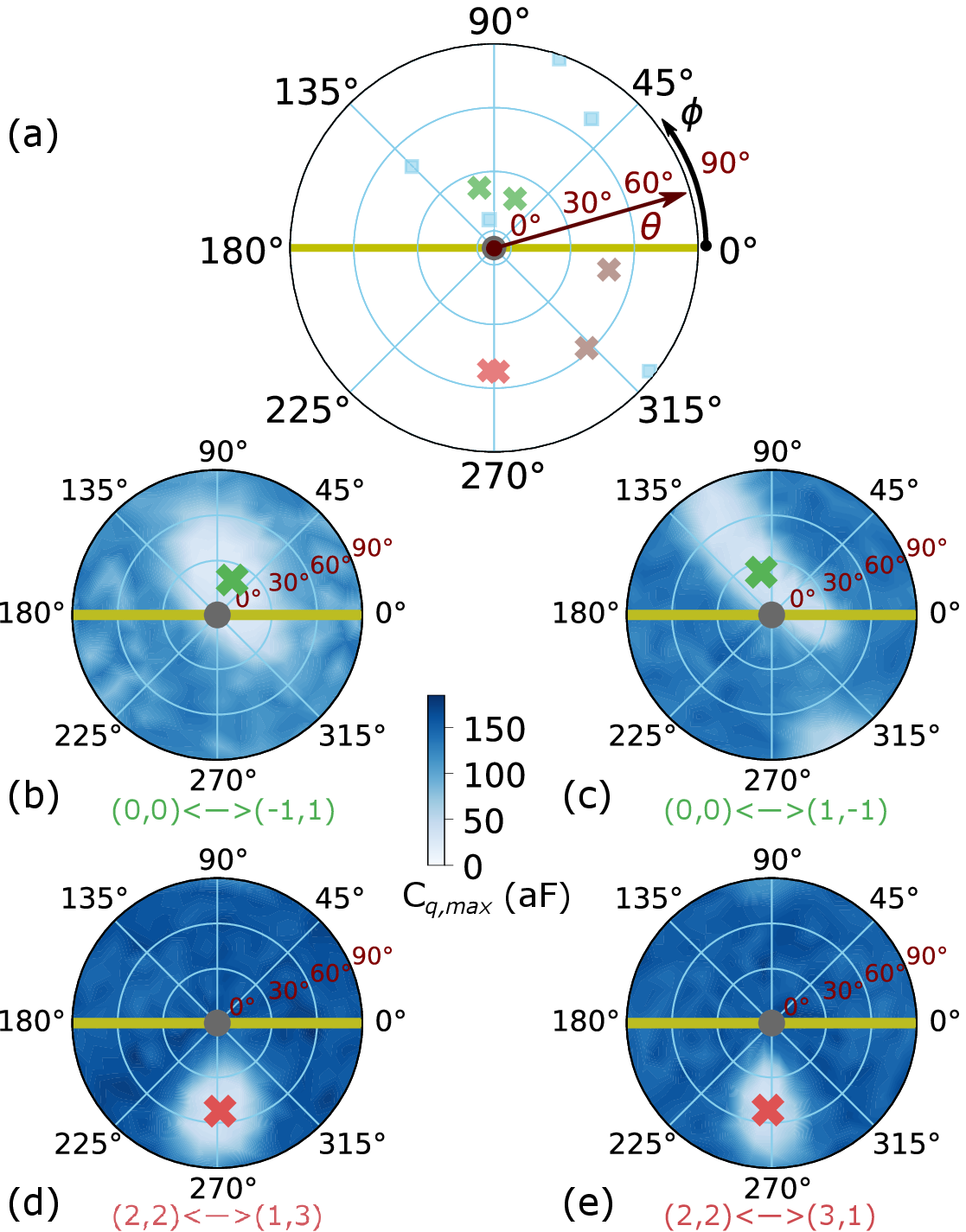}
\caption{\label{fig:Fig4} 
(a) Summary of the extracted $\vec{B}_{SO}$ directions for the even-occupied ICTs under study, with $|\vec{B}| = 50$~mT.
Crosses in the same color correspond to the ICT pairs with the same valence orbital. 
Squares mark the ICTs when the other ICT from a pair is not measured.
(b-e) Examples of the external magnetic field angle map of $C_{q,max}$ ($0^\circ \leq \theta \leq 90^\circ$), where the ICTs in (b,c) are within 2 charge transitions of those in (d,e).
These ICTs are labeled with the corresponding charge occupations ($n'_L,n'_R$) with respect to even charge numbers ($N'_L,N'_R$).
}
\end{figure}
%-------------------------------------------------Figure-4: end

% ================== 6. Orbitals variation ==================

        % 11- what is data in supplB
After analyzing an individual ICT, we look into the $\vec{B}_{SO}$ orientation of clusters of ICTs.
We rotate $\vec{B}$ while measuring $C_{q,max}$ for 16 neighboring ICTs (Suppl. C), where for odd-occupied ICTs, the extracted $C_{q,max}$ is independent of the $\vec{B}$ orientation.
On the contrary, the majority of even-occupied ICTs show a fairly well defined direction in which $C_{q,max}$ is suppressed, indicating the orientation of $\vec{B}_{SO}$. 
A few of the ICTs reveal a $C_{q,max}$ suppression in a peculiar pattern with no clear preferred direction, which will be discussed later.

        % 12- what is data in fig4
Fig.~\ref{fig:Fig4}(a) summarizes all of the extracted $\vec{B}_{SO}$ orientations, including some cases where we tune the barrier and plunger gates by a large amount. 
The crosses with the same color indicate pairs of ICTs of the same valence orbital.
The corresponding maps of extracted $C_{q,max}$ for two such ICT pairs are presented in Fig.~\ref{fig:Fig4}(b-e). 
Blue squares in (a) indicate the ICTs for which the other ICT from a pair is not measured. 
Because of the inversion symmetry shown in Fig.~\ref{fig:Fig2}(d-e), only the measurements for $0^\circ \leq \theta \leq 90^\circ$ are performed.

        % 13- what data means
The markers in Fig.~\ref{fig:Fig4}(a) show no preferred direction among the complete set of measured $\vec{B}_{SO}$ orientations. 
Notably, for a pair of ICTs corresponding to the same valence orbital, their $\vec{B}_{SO}$ orientations are much closer to each other than to other random pairs. 
This thereby support the hypothesis that the random orientation of $\vec{B}_{SO}$ arises from the complex shape of the electronic orbitals and the hard-to-predict local $\vec{E}$.
Imperfect alignment of the $\vec{B}_{SO}$ orientations within a pair of ICTs might be a consequence of a slight distortion of the confining potential, while the gates are tuning the dot occupancies.
The irregular shape of the $C_{q,max}$ suppression regions for some of the ICTs (e.g. Fig.~\ref{fig:Fig4}(c)) demonstrates that the description of SOI in terms of effective $\vec{B}_{SO}$ and isotropic g-factors is incomplete.
In Ref.~\cite{scherubl2019observation}, Sche\"ubl~et~al. discuss a topological nature of Weyl points between the lowest singlet and triplet states, which is equivalently manifested by the suppression of $C_q$ in our experiment. 
They show that the number of such Weyl points is not restricted to be 2 (like in Fig.~\ref{fig:Fig2}(d-e)), but may be 6 or even larger for rare cases.
The presence of more than 2 Weyl points might explain the highly irregular regions of $C_{q,max}$ suppression for some ICTs.

% ================== 7. Summary ==================

In summary, we study $\vec{B}_{SO}$ orientation using DGS in an InSb nanowire-based DQD.
At zero magnetic field, DGS can be employed as a charge parity meter, and ICTs with even total parity are identified for further $\vec{B}_{SO}$ characterization. 
When a finite external field $\vec{B}$ is rotated, the directions in which $C_{q,max}$ is suppressed reveal the $\vec{B}_{SO}$ orientation of even-occupied ICTs.
We model the dispersive signal at an even-occupied ICT, and find good agreement with the data for $\vec{B}\perp\vec{B}_{SO}$. 
However, for $\vec{B} \parallel \vec{B}_{SO}$, our model lacks a description of the suppressed spin relaxation rates due to Pauli blockade.
Finally, we find that $\vec{B}_{SO}$ for the ICT pairs with the same valence orbital have similar orientations, while the $\vec{B}_{SO}$ orientation varies enormously between different orbitals.
Our work indicates that considerations about $\vec{B}_{SO}$ orientation based purely on device design may often not apply to quantum dot systems.
Resolving whether the randomness of $\vec{B}_{SO}$ orientation persists in quantum devices, either based on nanowires or two-dimensional electron gases, is essential to assess the viability of different materials for quantum computing with spins and Majorana zero modes.
Moreover, DGS is shown to be an effective tool in characterizing the $\vec{B}_{SO}$ orientation, especially when transport measurements are not applicable. 
This result also opens new prospects for systems applying DGS that feature integrated capabilities for qubit characterization and readout, while avoiding increasing the complexity in the chip design \cite{plugge2017majorana, smith2020dispersive, maman2020charge}.

% ================== 7. ACKNOWledgement ==================
\section{Acknowledgement}
We appreciate J. Koski for suggestions regarding to the experiment and the manuscript.
We thank K. Li for assisting nanowire deposition, 
F. Borsoi, N. van Loo and J. Wang for useful advice on fabrication,
A. Palyi and Z. Scherübl for fruitful discussions on the measurement results,
also M. Hornibrook and D.J. Reilly for providing the frequency multiplexing chips,
O.W.B. Benningshof, R.N. Schouten, and J.D. Mensingh for valuable technical assistance. 
This work has been supported by the Netherlands Organization for Scientific Research (NWO) and Microsoft Quantum Lab Delft.
FKM acknowledges support from NWO under Veni grant (VI.Veni.202.034).

% ================== Reference ==================
\bibliography{main}

\end{document}